\documentstyle[11pt,newpasp,twoside,epsf]{article}
\begin{document}
\title{Cepheid, Tully-Fisher and SNIa Distances.}
\author{T. Shanks}
\affil{Dept. of Physics, Univ. of Durham, South Road, Durham, England}
\author{P.D. Allen}
\affil{Dept. of Astrophysics, Univ. of Oxford, Keble Road, Oxford, England}
\author{F. Hoyle}
\affil{Dept. of Physics, Drexel Univ., Disque Hall 816, Philadelphia, USA}
\author{N.R. Tanvir}
\affil{Div. of Physics and Astronomy, Univ. of Herts., Hatfield, England}
\begin{abstract}
We first discuss why the uncomfortable fine-tuning of the parameters of
the $\Lambda$-CDM cosmological model provides continuing, strong
motivation to investigate Hubble's Constant. Then we review  evidence
from the HST Key Project that there is a significant scale error between
raw Cepheid and Tully-Fisher distances.  An analysis of mainly HST
Distance Scale Key Project data shows a correlation between host galaxy
metallicity and the rms scatter around the Cepheid P-L relation, which
may support a recent suggestion that the P-L metallicity dependence is
stronger than expected. If Cepheids do have a significant metallicity
dependence then the Tully-Fisher scale error increases and the distances
of the Virgo and Fornax clusters extend to more than 20Mpc, decreasing
the value of H$_0$. Finally, if the Cepheids have a metallicity
dependence then so do Type Ia Supernovae since the metallicity corrected
Cepheid distances to eight galaxies with SNIa would then suggest that the SNIa
peak luminosity is fainter in metal poor galaxies, with
important implications for SNIa estimates of q$_0$ as well as H$_0$.
\end{abstract}

\section{Status of the $\Lambda$-CDM Cosmology}

One major motivation for studying Hubble's Constant is the complicated nature of
the current standard model in cosmology, $\Lambda$-CDM. In this model, to order
of magnitude, $\Omega_{baryon}\approx \Omega_{CDM}\approx \Omega_{\Lambda}$ and
this seems unnatural. The coincidence between the CDM and Baryon densities
worried some authors (Peebles 1984, Shanks 1985) when CDM was first postulated. 
The coincidence between $\Omega_{\Lambda}$ and the others worried many more (eg
Dolgov, 1983, Peebles and Ratra, 1988 and Wetterich, 1988). These fine-tuning
problems of the standard model are compounded by the fact that the inflation
model on which the standard model sits, was partly based on a fine-tuning
argument, the flatness-problem; to begin by eliminating one fine tuning problem
only to end up with several gives the appearance of circular reasoning!

Shanks (1985, 1991, 1999, 2001) noted that a simpler  model immediately
became available if H$_0$ actually lay below 50 kms$^{-1}$ Mpc$^{-1}$.
An inflationary model with $\Omega_{baryon}$=1 is then better placed to
escape the baryon nucleosynthesis constraint. Simultaneously, the low
value of H$_0$ means that the X-ray gas in the Coma cluster increases
towards the Coma virial mass and the lifetime of an Einstein-de Sitter
Universe extends to become compatible with the ages of the oldest stars.
Now this model does predict a first acoustic peak in the CMB anisotropy
at around l$\approx$330 which disagrees with the position of the first
peak at l=220$\pm$10 found  by the Boomerang experiment (Netterfield et
al, 2002). However, the above fine-tuning problems of the $\Lambda$-CDM model plus
the historical tendency for later data to overturn early confirmations
of  previous `concordance' models such as the isocurvature model in the
early 1980's and the SCDM model in the early 1990's suggests that  it
may be best, (a) to wait for MAP to confirm the Boomerang results before
abandoning other models and (b) to continue to study Hubble's Constant.

\begin{figure}
\plotfiddle{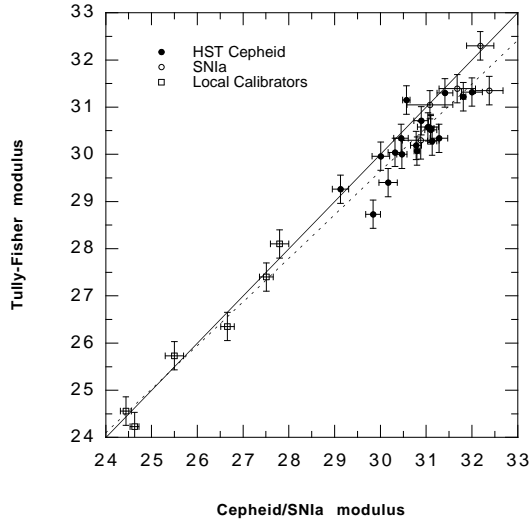}{2.5in}{0}{40}{40}{-120}{-65}
\caption{ A comparison between HST Cepheid and TF distances which suggests that
TF distances show a significant scale-error with the TF distance to galaxies at the 
distance of the Virgo cluster being underestimated by 22$\pm$5.2\%. The dashed line
shows the best fit with $(m-M)_{TF}= 0.915\pm0.036\times(m-M)_{Ceph}+2.204$.} 
\end{figure}

\section{A New Era for Determining H$_0$}
Some 25 galaxies have had Cepheids detected by HST. Seventeen of these were
observed by the HST Distance Scale Key Project (Freedman et al, 1994). Seven were
observed in galaxies with SNIa by Sandage and collaborators (eg Sandage et al,
1996) and M96 in the Leo I Group was observed by Tanvir et al (1995). In Fig. 1
we use these data to update the comparison of I-band TF distances of Pierce \&
Tully (1992) with HST Cepheid distances. As can be seen, the result implies that
TF distance moduli at Virgo underestimated by $\approx$22$\pm$5\%. This reduces
Tully-Fisher estimates of H$_0$ from $\approx$85 to
$\approx$65kms$^{-1}$Mpc$^{-1}$ (Giovanelli et al, 1997, Shanks 1997, Shanks,
1999, Sakai et al, 1999). The correlation of Cepheid residuals with line-width
suggests  TF distances may be Malmquist biased - possibly implying a bigger TF
scale error at larger distances. This clear problem for TF distances, which previously
has been the `gold standard' of secondary distance indicators, warns that errors
in the extragalactic distance scale may still be  seriously underestimated!

\begin{figure}
\plotfiddle{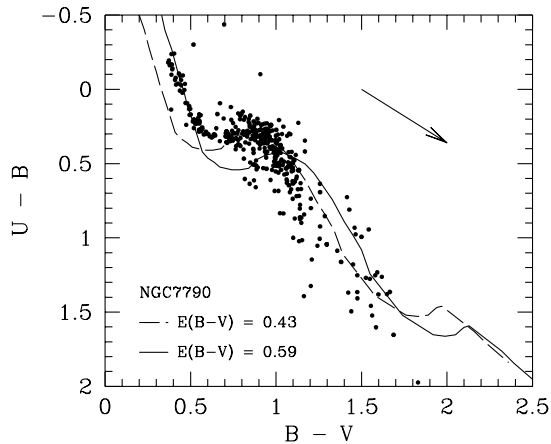}{2.5in}{0}{40}{40}{-120}{-65}
\caption{The UBV 2-colour plot from Hoyle, Shanks and Tanvir (2001) for the
Galactic open cluster, NGC7790, which contains 3 Cepheids. The data is not well
fitted by a solar metallicity, zero-age main sequence 2-colour diagram since
different values of the reddening, E(B-V) are implied by the B stars and the F
stars as shown. One interpretation is that the correct E(B-V) is given by the B
stars and that the F stars are showing UV-excess caused by the cluster
metallicity being significantly lower than Solar.}
\end{figure}

\section{NGC7790 Cepheid Metallicity  Dependence?}
New JKT 1.0m + CTIO 0.9m + UKIRT UBVK photometry of Cepheid Open Clusters by
Hoyle et al. (2001) has uncovered an anomaly in the NGC7790 UBV 2-colour
diagram, in that the F stars in the cluster show a strong  UV excess with respect
to zero-age main sequence stars (see Fig. 2). The result is confirmed by
independent photometric data (Fry, 1997, Fry and Carney, 1997) as shown in Fig. 7b
of Hoyle et al (2001). If the UV excess is caused by metallicity then NGC7790 would
have [Fe/H]$\approx$ -1.5 ! To keep the Galactic Cepheid P-L relation as tight
as previously observed implies that Cepheids may have a stronger metallicity
dependence, $\Delta M  \approx -0.66 \Delta[Fe/H]$, than previously expected, in
the sense that low metallicity Cepheids are intrinsically fainter. Currently we
are obtaining metallicities for the F stars in NGC7790 in order to distinguish this
interpretation from other explanations such as non-standard dust reddening towards 
this open cluster.

\begin{figure}
\plotfiddle{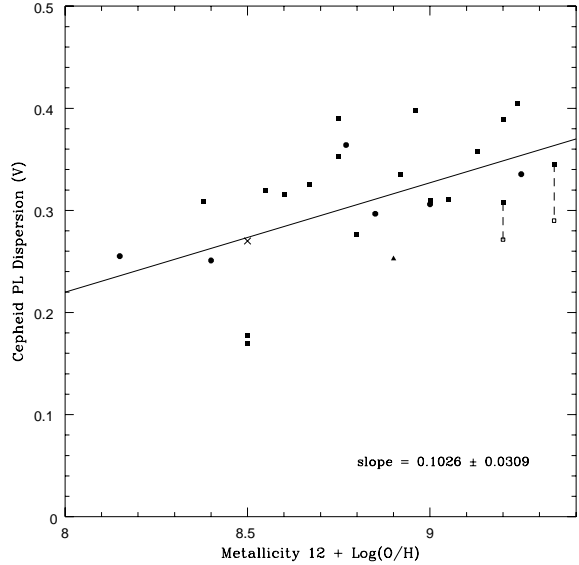}{2.5in}{0}{40}{40}{-120}{-75}
\caption{Relationship between mean r.m.s. dispersion (V-band) about the Cepheid
P-L relation and metallicity of HII regions in the vicinity of the Cepheids. Data
from the H$_{0}$ Key Project team is shown as squares. Sandage et al data is
shown as circles and the Tanvir et al data as a triangle. Finally, the LMC is
shown as a cross. Dashed vertical lines show the effect of removing outliers in
two P-L relations. The result of a least squares fit to the data
is also shown. }
\end{figure}

\section{HST Cepheid Metallicity Dependence}
Meanwhile, Allen \& Shanks (2001) have found an $\approx3\sigma$ correlation
between dispersion around the Cepheid P-L relation and galaxy metallicity for HST
Cepheid galaxies (see Fig. 3). This again suggests that the dependence of the Cepheid P-L
relation on metallicity may be more complex than previously expected.

\begin{figure}
\plotfiddle{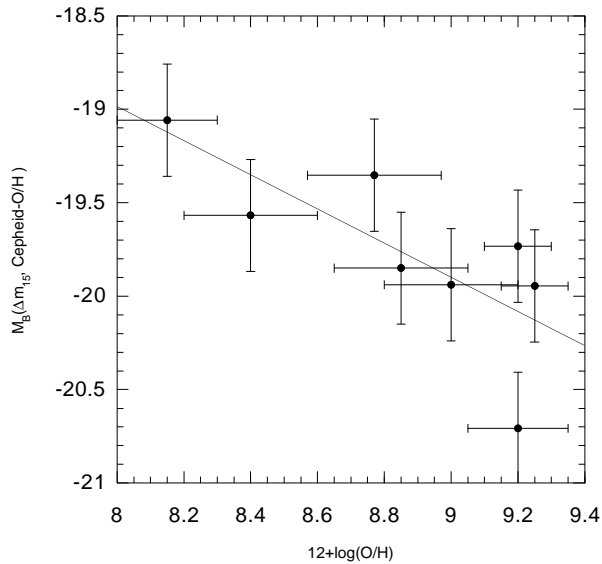}{2.5in}{0}{60}{60}{-180}{-215}
\caption{ The SNIa absolute magnitude-metallicity relation using the SNIa
peak magnitudes of Gibson et al. (1999), now corrected for $\Delta m_{15}$  and
Cepheid metallicity. The least-squares fitted line,  $M_B=-0.92(\pm0.33)logW-11.65$, 
is also shown.}
\end{figure}

Allen \& Shanks (2001) also obtain Cepheid distances  via truncated maximum
likelihood P-L fits to account for magnitude incompleteness caused by the above,
non-negligible, dispersion in the HST P-L relations. They found that Cepheid galaxy
distances at the limit of HST reach are too low. The higher than expected  P-L
dispersion  for distant, metal-rich galaxies accentuates this effect. The
conclusion is that current HST Cepheid distance moduli may be underestimated  by
more than 0.5 mag at the redshift of Virgo and Fornax due to  both metallicity
and statistical incompleteness bias. The TF distances discussed above are then
underestimates by approximately 1 magnitude.

Eight HST Cepheid galaxies also have Type Ia distances. Correcting the Cepheid
scale for metallicity and incompleteness bias as above and then using these
distances to derive peak luminosities using the SNIa data from Gibson et al (2000),
implies a strong correlation between Type Ia peak luminosity and metallicity.
Such a scatter in SNIa luminosities could easily be disguised by magnitude
selection effects at moderate redshifts. At higher redshift the
correlation is in the right direction to explain away the need for a cosmological
constant in the Supernova Hubble Diagram results, since galaxies at high redshift
might be expected to have lower metallicity. Thus the conclusion is that if
Cepheids have a strong metallicity dependence then so have SNIa and therefore SNIa
estimates of q$_0$ and H$_0$ may require  significant corrections for
metallicity.

\section{Conclusions - Implications for H$_0$ and SNIa Cosmology}

Our conclusions are as follows:-

\begin{itemize}

\item Key Project  HST Cepheid distances imply Tully-Fisher distances at Virgo/Fornax are
underestimated by $\approx22\pm5$\%, reducing H$_0$ from $\approx$85 to
$\approx$65kms$^{-1}$Mpc$^{-1}$.

\item TF distances may be Malmquist biased,  suggesting there may be a 
bigger TF scale error at larger distances.

\item If the UV excess of F stars in open cluster NGC7790 is caused by low
metallicity then  Cepheids have  a strong metallicity dependence,  $\Delta M 
\approx -0.66  \Delta[Fe/H]$.

\item  Current HST Cepheid distances may be significantly underestimated at
Virgo/Fornax redshifts due to metallicity and magnitude incompleteness  bias, implying that
values of H$_0<$50kms$^{-1}$Mpc$^{-1}$  may still not be ruled out.

\item If Cepheids have a strong metallicity dependence then so have SNIa . Thus 
significant metallicity corrections may need to be applied to the Type Ia Hubble 
Diagram before reliable estimates of q$_0$ or H$_0$ can be made.

\end{itemize}

%\acknowledgments I am grateful to P.D.Allen, (Univ. of Oxford), F.Hoyle (Drexel University) and
%N.R. Tanvir (University of Hertfordshire) for help with the research reported in this paper.

\end{document}